\documentclass[%
 aip,
 amsmath,amssymb,
preprint,%
]{revtex4-1}

\usepackage{graphicx}
\usepackage{dcolumn}
\usepackage{braket}
\usepackage{bm}

\usepackage[utf8]{inputenc}
\usepackage[T1]{fontenc}
\usepackage{mathptmx}
\usepackage{etoolbox}

\makeatletter
\def\@email#1#2{%
 \endgroup
 \patchcmd{\titleblock@produce}
  {\frontmatter@RRAPformat}
  {\frontmatter@RRAPformat{\produce@RRAP{*#1\href{mailto:#2}{#2}}}\frontmatter@RRAPformat}
  {}{}
}%
\makeatother
\begin{document}

\preprint{}

\title[]{Thermal biphotons}
\author{Ohad Lib}
\author{Yaron Bromberg}%
 \email{Yaron.Bromberg@mail.huji.ac.il}
\affiliation{Racah Institute of Physics, The Hebrew University of Jerusalem, Jerusalem, 91904 Israel
\\This article may be downloaded for personal use only. Any other use requires prior permission of the author and AIP Publishing. This article appeared in APL Photonics 7, 031301 (2022) and may be found at https://doi.org/10.1063/5.0085342.}%

\date{\today}

\begin{abstract}
The observation of the Hanbury Brown and Twiss (HBT) effect with thermal light marked the birth of quantum optics. All the thermal sources considered to date did not feature quantum signatures of light, as they consisted of independent emitters that emit uncorrelated photons. Here, we propose and demonstrate an incoherent light source based on phase-randomized spatially entangled photons, which we coin \textit{thermal biphotons}. We show that in contrast to thermal light, the width of the HBT peak for thermal biphotons is determined by their correlations, leading to violation of the Siegert relation and breakdown of the speckle-fluctuations interpretation. We further provide an alternative interpretation of the results by drawing a connection between the HBT effect and coherent backscattering of light. Finally, we discuss the role of spatial entanglement in the observed results, deriving a relation between the Schmidt number and the degree of violation of the Siegert relation under the double-Gaussian approximation of spontaneous parametric down conversion (SPDC). Our work reflects new insights on the coherence properties of thermal light in the presence of entanglement, paving the way for entanglement certification using disorder averaged measurements.
\end{abstract}

\maketitle

One of the most important steps towards our modern understanding of optical coherence was the observation of the Hanbury Brown and Twiss (HBT) effect, in which the size of a thermal light source was inferred using intensity correlation measurements between two detectors\cite{brown1956test,brown1956correlation}. Although being completely consistent with the classical theory of light, the observation of the HBT effect triggered a heated debate on the quantum nature of photons\cite{brannen1956question}. The full quantum description of the HBT effect was finally given by Glauber in his seminal work on the quantum theory of optical coherence\cite{glauber1963photon,glauber1963quantum}, marking the birth of the field of quantum optics. Following these results, second-order coherence (or intensity interferometry) has gained significant importance in quantum optics, playing a major role in phenomena such as Hong-Ou-Mandel interference\cite{hong1987measurement} and the characterization of single and entangled photon sources\cite{kwiat1995new,kurtsiefer2000stable}.

\begin{figure}[h!]
 \centering
 \includegraphics[width=0.5\textwidth]{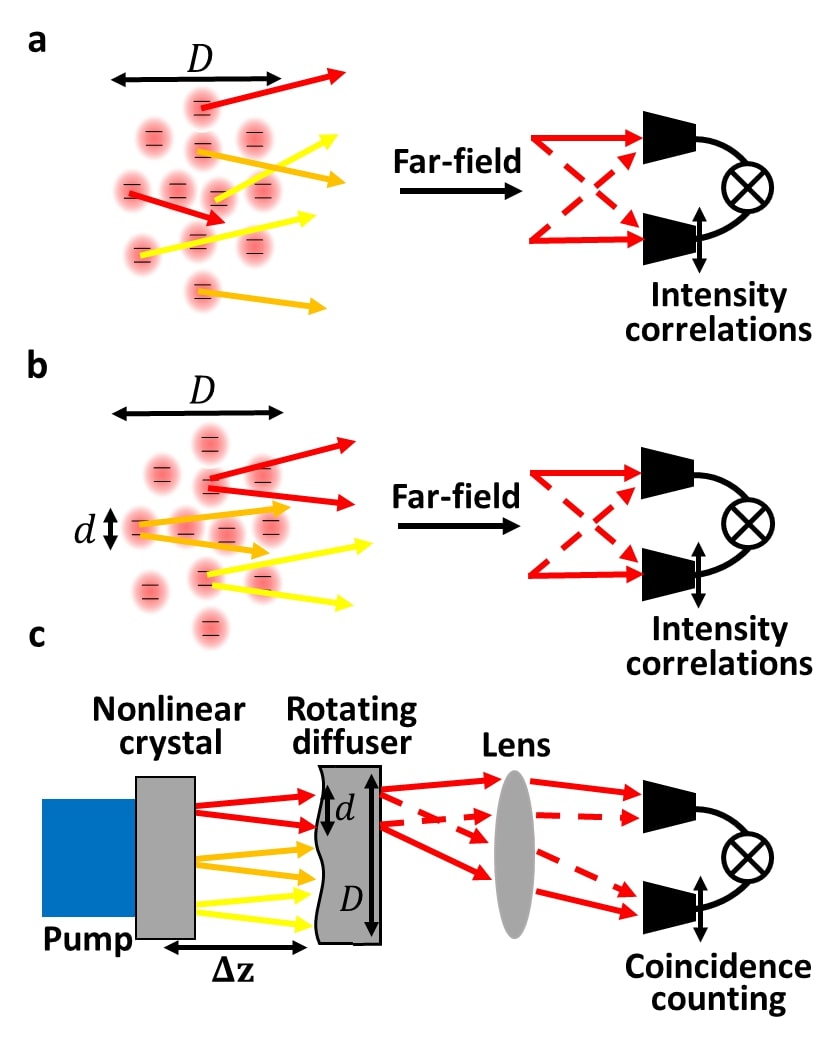}%
 \caption{\label{fig1}(a) Illustration of an HBT measurement with a thermal source of transverse width $D$. Photons are emitted by uncorrelated emitters distributed over the entire source. (b) Illustration of an HBT measurement with thermal biphotons. Pairs of photons are emitted by two-photon emitters of transverse size $d$. (c) Experimental implementation of a pseudo-thermal biphoton source. Pairs of correlated photons generated by spontaneous parametric down conversion are scattered by a rotating diffuser.}
 \end{figure}

In all demonstrators to date of the spatial HBT effect, the sources were spatially incoherent thermal sources, which are modeled by a large number of independent emitters that emit uncorrelated photons (Fig.\ref{fig1}(a))\cite{goodman2015statistical}. For such sources, the classical interpretation of fluctuating Gaussian fields, and the quantum interpretation of photon bunching, predict exactly the same features of the HBT effect: the second order coherence exhibits a 2:1 peak-to-background ratio with a width that is inversely proportional to the source size. Similarly, it is generally believed that the Gaussian fields description of thermal ghost imaging yields equivalent results to the quantum theory, up to a contrast difference\cite{gatti2004ghost,scarcelli2006can,gatti2007comment,shapiro2008computational,erkmen2008unified,bromberg2009ghost}. An intriguing question is whether the equivalence between the Gaussian field and the photon bunching interpretations holds for other types of incoherent sources. To study this question, we discuss a class of spatially incoherent sources, coined thermal biphotons, that are made of independent emitters that emit \textit{pairs} of photons. Beyond reflecting new insights on the classical and quantum description of spatial coherence in the presence of two-photon correlations, understanding the coherence properties of thermal biphotons may open the door for new applications of two-photon sources with low spatial coherence.

In this work, we theoretically and experimentally study the coherence properties of thermal biphotons. We show that thermal biphotons exhibit a 2:1 HBT peak, as in the standard HBT effect, indicating their thermal nature. The width of the HBT peak, however, is different for thermal biphotons. We show that in contrast to the standard HBT effect with thermal light, for thermal biphotons the quantum and the classical fluctuating-speckle pictures are not equivalent. They predict different widths for the HBT peak, leading to violation of the Siegert relation and to the breakdown of the fluctuating-speckle interpretation of the HBT effect. We give an alternative interpretation of the observed results by unveiling a relation between the HBT effect and coherent backscattering of light (CBS) in complex media. Finally, we discuss and highlight the role of spatial entanglement in our results, focusing on the case of entangled pairs generated via spontaneous parametric down conversion (SPDC) under the double-Gaussian approximation.

We begin by briefly describing the standard HBT effect with thermal light. Light from a thermal source of size $D$ is measured at the far-field using two detectors at angular separation $\Delta\theta$ relative to the source (Fig.\ref{fig1}(a)). Since the thermal light field has Gaussian statistics, the instantaneous intensity distribution at the detectors' plane exhibits rapidly fluctuating speckle patterns\cite{goodman2007speckle,aspect2020hanburry}. If the two detectors lie within the same speckle grain, their intensities fluctuate correlatively, and $g^{(2)}(\Delta \theta)\approx \frac{\braket{I^{2}}}{\braket{I}^{2}}=2$, where the last equality results from the exponential statistics of thermal light\cite{scully1999quantum}. Otherwise, the detectors measure intensities at uncorrelated speckle grains, yielding $g^{(2)}(\Delta \theta)=1$. The width of the 2:1 peak in $g^{(2)}(\Delta\theta)$ is thus determined by the speckle grain size, which scales like $\lambda/D$, where $\lambda$ is the wavelength of the thermal light. An equivalent classical view of the HBT effect is given by utilizing the Siegert relation, $g^{(2)}(\Delta\theta)=1+|g^{(1)}(\Delta\theta)|^{2}$, which connects the first- and second-order coherence functions of thermal light\cite{siegert1943fluctuations}. According to the Van Cittert-Zernike theorem\cite{goodman2015statistical}, the width of the first-order coherence function $g^{(1)}(\Delta\theta)$ is determined by the Fourier transform of the intensity of the thermal source. Therefore, using Siegert's relation, one again finds that the width of the HBT peak is the same, and scales as $\lambda/D$. The quantum description of the thermal HBT effect yields the exact same results as the above classical interpretation. In the framework of quantum mechanics, one has to sum all different two-photon amplitudes leading to a detection of a photon in each detector\cite{fano1961quantum}. The paths corresponding to two such amplitudes are illustrated in fig.\ref{fig1}(a). Indeed, if the distance between the detectors is small enough, the different amplitudes add coherently, leading to an increase in intensity correlations.

Next, we consider the case of thermal biphotons, in which a source of size $D$ is comprised of two-photon emitters, each having a characteristic size of $d$ (Fig.\ref{fig1}(b)). We will show that for thermal biphotons, the width of the HBT peak is $D/d$ times wider than the HBT peak for thermal light, while the width of the first-order coherence function remains unchanged. To show this, inspired by the experimental implementation described below, we model a source of thermal biphotons using a quantum state of the form $ \ket{\psi}  = \int dr_s dr_i \sqrt{P(r_s,r_i)} exp(i( \Phi (r_s,r_i )))exp(i( \phi (r_s)+ \phi (r_i))) \ket{ 1_{r_s},1_{r_i}}$, where $\ket{ 1_{r_s},1_{r_i}}$ is a Fock state describing the emission of two photons (coined signal and idler photons) at positions $r_s$ and $r_i$. $\sqrt{P(r_s,r_i )} exp(i(\Phi(r_s,r_i )))$ is a general polar representation of the distribution of the emitted photons, which is symmetric to the exchange of the two photons, and the phase of different two-photon emission $\phi (r_s)+ \phi (r_i)$ is assumed to randomly fluctuate in space and time.

As the source we consider describes the emission of pairs of photons according to the spatial distribution given by the function $P(r_s,r_i)$, it is useful to define the size of a two-photon emitter as the typical distance between two correlatively emitted photons, and the width $D$ of the source as the width of the single-photon emissions distribution. We therefore quantitatively define the characteristic size of the two-photon emitters as the width of the correlations between the photons at the plane of the source, $d^2 \equiv \int dr_s dr_i(r_s-r_i)^2 P(r_s,r_i)$, yielding that $d$ is $\sqrt{2}$ times the standard deviation of the two-photon distance distribution $h_- (r_- )\equiv\int dr_+ P(r_s,r_i)$, where $r_\pm =(r_i\pm r_s)/\sqrt{2}$ (see Supplementary information). Similarly, the characteristic size of the source, $D$, is defined as the standard deviation of the marginal single-photon distribution $h_s (r_s )\equiv \int dr_i P(r_s,r_i )=\bra{\psi} a_s^{\dagger} (r_s ) a_s (r_s )\ket{\psi} $, where $a_s^\dagger  (r_s )$ is the creation operator of a signal photon at position $r_s$ at the plane of the source.

Using this model, we can now derive the expression for the second-order coherence function of thermal biphotons (see Supplementary information):

\begin{equation} \label{eq1}
 g^{(2)}(\Delta\theta) =\frac{\bra{\psi} a^{\dagger}_s(\theta_1) a^{\dagger}_i(\theta_2) a_i(\theta_2) a_s(\theta_1)\ket{\psi}}{\bra{\psi} a_s^{\dagger}(\theta_1) a_s(\theta_1) \ket{\psi} \bra{\psi} a_i^{\dagger}(\theta_2) a_i(\theta_2)\ket{\psi}} 
 = 1+\widetilde{h_-} (\sqrt{2} k\Delta \theta)
\end{equation}

where $k$ is the wavenumber, $a_s (\theta) = \sqrt{(k/2 \pi)} \int dr_s a_s (r_s )exp(-ik \theta r_s )$ is the annihilation operator for a signal photon with (paraxial) transverse wavevector $q=k\theta$ at the plane of the source and $\widetilde{h_-} (k\Delta \theta)$ is the Fourier transform of the two-photon distance distribution $h_- (r_-)$. The first-order coherence function can be evaluated as well, yielding $g^{(1)} (\Delta \theta) = \widetilde{h_s} (k \Delta \theta)$, where $\widetilde{h_s} (k \Delta \theta)$ is the Fourier transform of $h_s (r_s)$, in accordance with the Van Cittert-Zernike theorem. For a separable quantum state, corresponding to a standard thermal source without correlations, $P(r_s,r_i)$ can be written as $P(r_s,r_i )=P_1 (r_s ) P_1 (r_i)$. In this case, Eq.(\ref{eq1}) captures the standard thermal HBT effect, namely, the Siegert relation holds $g^{(2)}(\Delta\theta)=1+|\widetilde{h_s} (k \Delta \theta)|^{2}$. Thus for separable states, the width of $g^{(2)}(\Delta\theta)$ is proportional to the width of $\widetilde{h_s} (k \Delta \theta)$, and $\Delta \theta_{HBT} \propto \lambda/D$.

For thermal biphotons, described by a non-separable state, the photons exhibit spatial correlations, such that the size of the two-photon emitters is smaller than the total size of the source ($d<D$). As the width of $h_- (r_-)$ is determined by $d$, the width of the HBT peak corresponds to $\Delta \theta_{HBT}\propto \lambda/d$ for thermal biphotons, yielding an HBT peak which is $D/d$ times wider relative to thermal light. In contrast, the width of the first-order coherence function is still determined by the width of the source, $D$, and is not affected by the size of the two-photon emitters. These results show that in the case of thermal biphotons, the classical picture of a fluctuating speckle pattern for the HBT effect collapses, and the Siegert relation is broken.

To experimentally study the coherence features of thermal biphotons, we realize thermal biphotons using scattering of spatially entangled photons by a rotating diffuser. Scattering of spatially entangled photons exhibits a wide range of effects, such as two-photon speckle patterns\cite{beenakker2009two,peeters2010observation,cande2013quantum,klein2016speckle,lib2020real} and the observation of bosonic, fermionic, and anyonic symmetries\cite{van2012bosonic}. Here, in analogy with the creation of pseudo-thermal light by a laser beam and a rotating diffuser\cite{arecchi1965measurement}, we utilize scattering of spontaneous parametric down converted light to create a source of pseudo-thermal biphotons and study their coherence properties. This is in contrast to previous studies that considered coherence properties of entangled photons produced using either a coherent or an incoherent pump beam, yet without dynamic scattering\cite{joobeur1996coherence,saleh2000duality,defienne2019spatially,zhang2019influence}.

A simplified illustration of the experimental setup is presented in fig.\ref{fig1}(c). Spatially entangled photons are created via type 1 spontaneous parametric down conversion (SPDC), by pumping an $8\ mm$ long nonlinear Beta Barium Borate (BBO) crystal with a continuous-wave pump beam ($\lambda=404\ nm$). The twin photons are phase-randomized by passing through a rotating diffuser. The second-order coherence function is measured at the far-field using two single-photon detectors and a coincidence counting circuit (Swabian Time Tagger). In the experiment, one detector is always kept stationary at $\theta_i=0$ while the other scans the angle $\theta_s$. A polarizer and $80\ nm$ bandpass filters (not shown) are used to select the wavelength and polarization of the measured photons. The width of the two-photon emitter, $d$, which in our setup is determined by the distance between the signal and idler photons at the plane of the diffuser, can be tuned by changing the distance $\Delta z$ between the nonlinear crystal and the rotating diffuser.

\begin{figure}[h!]
 \centering
 \includegraphics[width=0.5\textwidth]{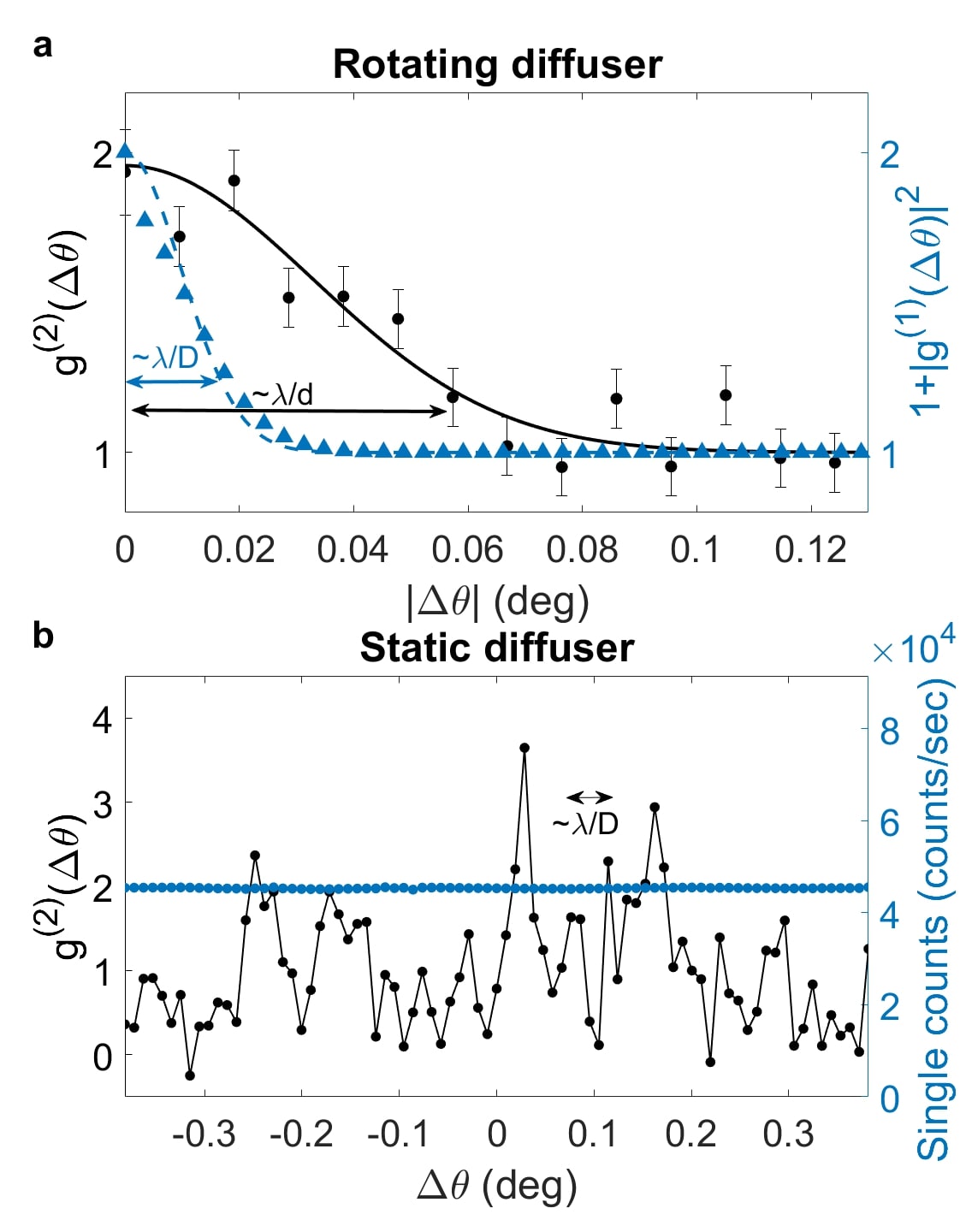}%
 \caption{\label{fig2}(a) Second-order $g^{(2)}(\Delta \theta)$ and first-order $g^{(1)}(\Delta \theta)$ coherence measurements of thermal biphotons. The data is taken with $\Delta z=35\ mm$ ($D\approx 2d$). We use the double-Gaussian approximation of SPDC and apply it to the theory of thermal biphotons to derive an explicit and simple model for the results\cite{law2004analysis,walborn2010spatial}. We obtain the following approximations for the coherence functions of thermal biphotons, $g_{DG}^{(2)} (\Delta \theta )=1+exp(-(d^2 (k\Delta \theta )^2)/2)$ and $|g_{DG}^{(1)} (\Delta \theta )|^2=exp (-D^2 (k\Delta \theta )^2 )$ (see Supplementary information). We apply this model for the observed results for the first- and second-order coherence functions (dashed and solid curves, respectively), yielding $d=230 \pm 16 \mu m$ and $D=525 \pm 2 \mu m$. (b) The $g^{(2)}(\Delta \theta)$ measurement with a static diffuser at $\Delta z=35\ mm$ ($D\approx 2d$, black) exhibits a two-photon speckle with a typical speckle size of $\lambda/D$. The single photon counts (blue) are homogeneously distributed and do not exhibit a speckle pattern. In all $g^{(2)}(\Delta \theta)$ measurements, accidental counts are subtracted.}
 \end{figure}

In the experiment, we can tune the size of the two-photon emitters such that $D\approx 2d$, by setting $\Delta z=35\ mm$. The second-order coherence function $g^{(2)}(\Delta \theta)$ exhibits an 2:1 HBT peak (Fig.\ref{fig2}(a), black dots), yet its width is determined by the size of the two-photon emitters rather than by the total width of the source. To study the relation between the first- and second-order coherence functions, we measure the intensity distribution of the source at the diffuser's plane using a CMOS camera, and compute its Fourier transform to obtain $g^{(1)}(\Delta\theta)$ according to the Van Cittert-Zernike theorem \cite{goodman2015statistical}. We thus observe $1+|g^{(1)}(\Delta\theta)|^2$ (Fig.\ref{fig2}(a), blue triangles). A clear violation of Siegert's relation is observed, as the widths of the two peaks are significantly different, supporting the theoretical prediction of Eq.(\ref{eq1}).

Thus far, we have shown that the width of the HBT peak for thermal biphotons is determined by the size of the two-photon emitter $d$ rather than by the size of the entire source $D$, i.e. $\Delta \theta_{HBT}\propto\lambda/d$. From the classical speckle interpretation of the HBT effect with thermal light, where the width of the HBT peak is equal to the width of a speckle grain $\Delta \theta_{speckle}\propto \lambda/D$, one might expect that for thermal biphotons the width of the HBT peak will be equal to the width of the two-photon speckle grain\cite{van2012bosonic}. We thus examine the two-photon speckle pattern obtained by measuring the second-order coherence function for a static diffuser (Fig.\ref{fig2}(b)). Since in SPDC the single photon distribution is spatially incoherent, no speckle pattern is observed in the intensity measurement, even though the diffuser is static (blue dots)\cite{peeters2010observation,lib2020real}. Nevertheless, the two-photon wavefunction is spatially coherent, thus exhibiting a two-photon speckle pattern (black dots). The width of a two-photon speckle grain is $\Delta \theta_{speckle}\propto\lambda/D$\cite{peeters2010observation}, which in contrast to the standard HBT effect, is not the same as the width of the HBT peak. Thus, we show that in contrast to the standard HBT effect, for thermal biphotons the quantum and the classical speckle-based interpretations are not equivalent, as the width of the HBT peak is different from the speckle size.

\begin{figure}[h!]
\centering
 \includegraphics[width=0.7\textwidth]{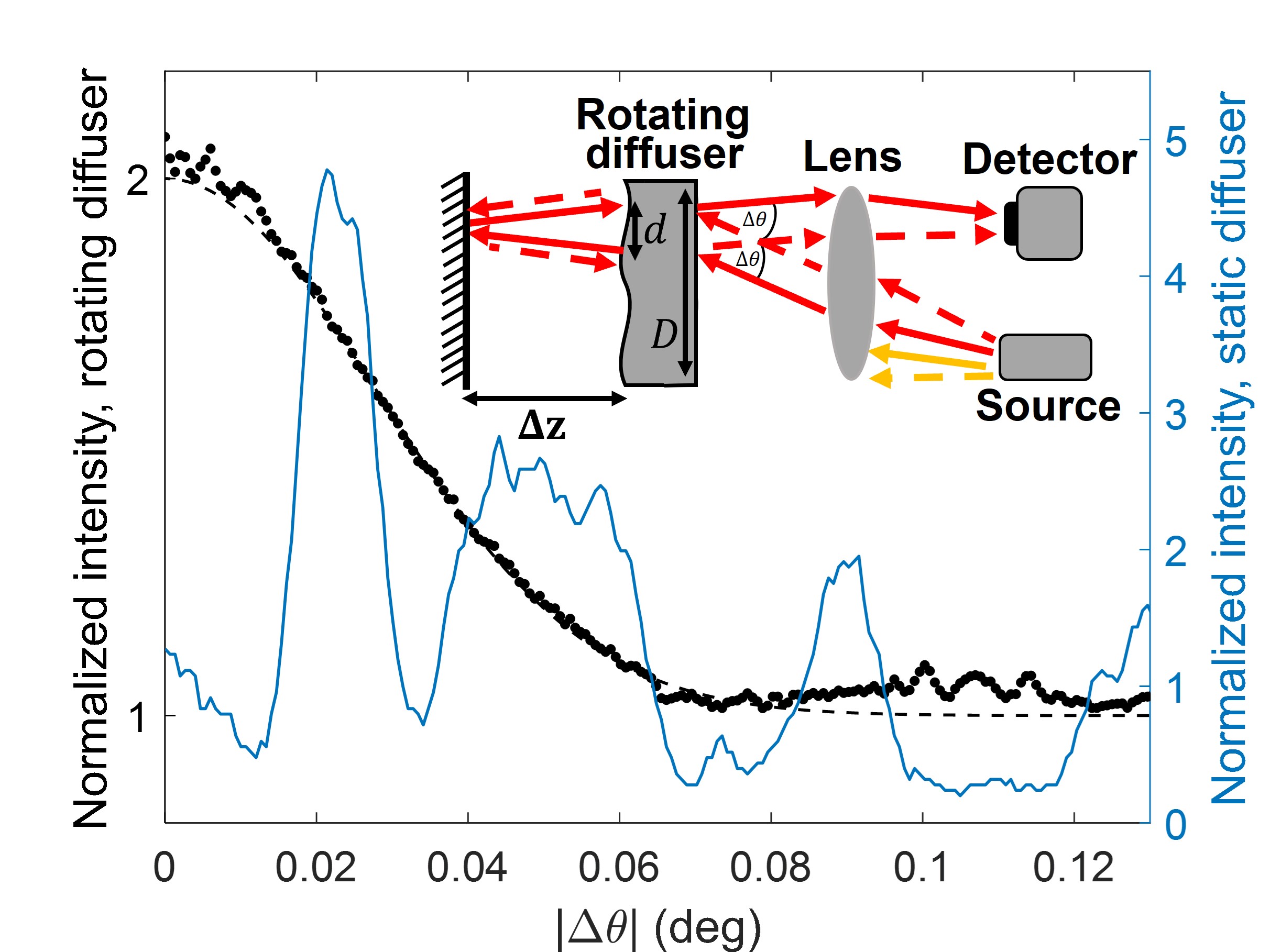}%
 \caption{\label{fig4} Illustration of the advanced wave picture setup is presented in the inset. One of the single-photon detectors is replaced with a laser, illuminating a diffuser-mirror-diffuser configuration. The intensity distribution is recorded at the far-field for $\Delta z=35 mm$, as a function of the angular separation between the incident and backscattered fields $\Delta \theta$ (see inset). An enhanced backscattering is observed for a measurement with a rotating diffuser (black dots), which is presented together with a theoretical fit to the double-Gaussian model. For a static diffuser, a speckle pattern is observed (blue line).}
 \end{figure}

While we have showed that the fluctuating-speckle interpretation fails to describe the coherence properties of thermal biphotons, we give an alternative classical interpretation that is surprisingly related to coherent backscattering of light (CBS) \cite{akkermans1986coherent}. By virtue of Klyshko's advanced wave picture, one of the single-photon detectors can be replaced with an equivalent classical source, the nonlinear crystal acts as a mirror, and the coincidence measurement can be replaced by measuring the intensity at the plane of the second detector\cite{klyshko1988awp}. Illustration of the equivalent advanced wave picture setup for our experiment is presented in the inset of fig.\ref{fig4}, where a classical, well-collimated, beam propagates through a diffuser-mirror-diffuser configuration, and measured at the far-field of the diffuser. This configuration exhibits coherent backscattering (CBS), in which an enhanced disorder-averaged intensity is observed in the backscattered direction, due to the constructive interference of reciprocal paths\cite{akkermans1986coherent,jakeman1988enhanced}. Two such paths are illustrated in the inset of fig.\ref{fig4} (solid and dashed red arrows). Each of the paths exits the effective scattering medium at the entry point of the other and are therefore reciprocal when measuring at $\Delta \theta = 0$. The mean transverse distance between the entry and exit points of the paths corresponds to an effective transport mean free path, given by $l = \braket{|r_{in}-r_{out}|}$, where $r_{in}$ and $r_{out}$ are the transverse positions of the entry and exit points respectively. The intensity in the backscattered direction relative to the background is twice as large, as in the HBT effect. The width of the CBS peak is determined by the effective transport mean free path, which is equivalent to the two-photon emitter size $d$, thus explaining the width of the HBT peak of thermal biphotons. To experimentally demonstrate the connection between thermal biphotons and coherent backscattering of a laser beam, we have measured the CBS peak obtained for $\Delta z=35 mm$ (Fig.\ref{fig4}, black). The curve is fitted according to the double-Gaussian approximation discussed above, yielding $d=264\pm 1 \mu m$. In addition, we measure the speckle pattern observed with a static diffuser in the advanced wave picture, demonstrating that the width of the CBS peak is indeed larger than the width of a speckle grain (Fig.\ref{fig4}, blue).

\begin{figure}[h!]
\centering
 \includegraphics[width=0.7\textwidth]{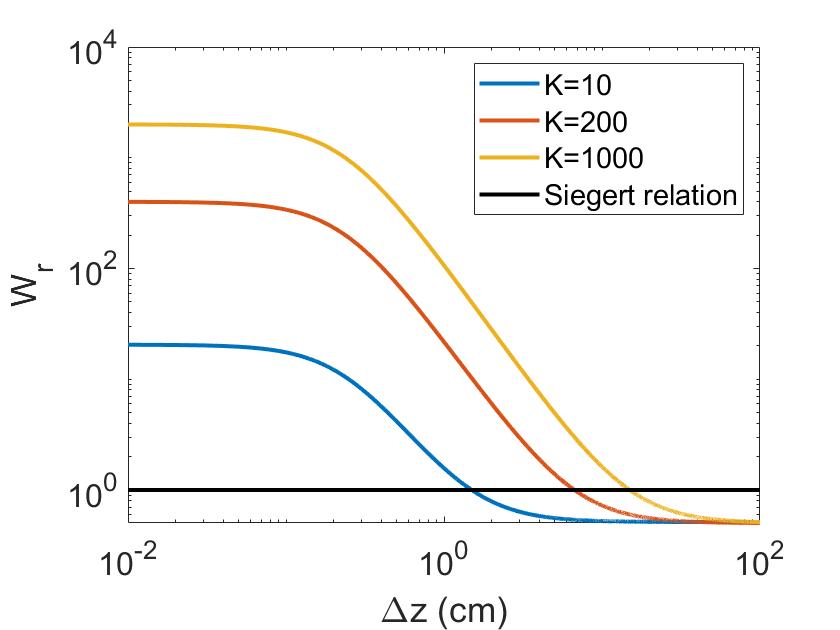}
 \caption{\label{fig5} The ratio between the area of the HBT peak and that of $|g^{(1)}|^2$ as a function of the distance between the nonlinear crystal and the rotating diffuser, under the double-Gaussian approximation (evaluated according to equation S18 in the supplementary material). The results are plotted for different Schmidt numbers obtained by varying the waist of the pump beam while keeping the length of the crystal fixed at $L=8mm$.}
 \end{figure}

Finally, we would like to discuss the role of spatial entanglement in the observed results. To this end, we consider three test cases of quantum states impinging onto the diffuser: (1) general pure states, (2) SPDC light under the double-Gaussian approximation and (3) mixed states with strong classical correlations. For pure states, entanglement is indeed a necessary condition for the violation of the Siegert relation, since, as discussed following Eq.(\ref{eq1}), separable pure states respect the Siegert relation. Therefore, under the assumption that the quantum state impinging onto the rotating diffuser is pure, the breakdown of the Siegert relation can be used to certify entanglement. Furthermore, for SPDC-generated photon pairs under the double-Gaussian approximation, we can relate the ratio between the area of the HBT peak and that of the first-order coherence function with the Schmidt number, $K$, which counts the effective number of entangled Schmidt modes\cite{law2004analysis}. Defining the ratio $W_r (\Delta z)=2D^2/d^2$, and looking at $\Delta z=0$ and $K \gg 1$, we obtain $W_r = 2(K+0.25)$, which relates the Schmidt number and the strength of the violation of the Siegert relation under the double-Gaussian approximation. For $\Delta z\neq 0$ and a general Schmidt number, an exact analytical relation between $W_r$ and the Schmidt number can be derived as well (supplementary information). The ratio $W_r$ as a function of $\Delta z$ is plotted in fig.\ref{fig5} for different Schmidt numbers. The violation of the Siegert relation is stronger for larger $K$ and for small $\Delta z$, where the spatial correlations are the strongest. It is interesting to note that for large enough $\Delta z$ the width of the HBT peak can be smaller than that of the first-order coherence function. 

While we have shown that entanglement is required for the violation of the Siegert relation for pure states, this is not generally the case for mixed states. For example, we consider a mixed state with strong (classical) spatial correlations at the plane of the diffuser, $\rho = \int dr P(r)\ket{2_{r}}\bra{2_{r}}$. While having only classical correlations, this state leads to the violation of the Siegert relation. Nevertheless, as is typically encountered in quantum interference effects, such tailored classical states can reproduce part of the effect, yet at the cost of altering other results. The best known example for such a case is Hong-Ou-Mandel (HOM) interference\cite{hong1987measurement}, where optimized classical states can exhibit an HOM dip, yet either with a lower visibility or with additional first-order interference that is not present in the quantum effect\cite{ghosh1987observation}. In the case of thermal biphotons, one must account for two key results. Namely, the violation of the Siegert relation (Fig.\ref{fig2}(a)), and the observation of a high-contrast speckle pattern with a smaller speckle width compared with that of the HBT peak (Fig.\ref{fig2}(b)). While classically correlated mixed states can break the Siegert relation, their reduced purity inhibits the observation of a high-contrast speckle pattern\cite{beenakker2009two}. Such states therefore cannot reproduce the results of thermal biphotons, and they do not break the fluctuating-speckle interpretation of HBT. 

In conclusion, we have discussed a source of thermal light, coined thermal biphotons, in which the photons are emitted in pairs from independent emitters. We have shown both theoretically and experimentally that in contrast to the standard HBT effect, the width of the HBT peak for thermal biphotons is determined by the spatial correlations between the photons rather than by the total size of the source. Therefore, we obtain that the width of the HBT peak can be wider than the speckle size. As a result, we observe a clear violation of the Siegert relation in the presence of correlations. We use the advanced wave picture interpretation of our experiment to demonstrate a connection between the HBT effect and coherent backscattering. Finally, we discuss the role of entanglement in the observed results, considering both pure and mixed states. We believe that further studies of this class of thermal light can lead to future applications. For example, inspired by classical imaging schemes that utilize pseudo-thermal light for structured illumination and speckle-free imaging\cite{lim2008wide,bromberg2009ghost,mudry2012structured,redding2013low,oh2013sub,pepe2017diffraction,gregory2020imaging,defienne2021polarization}, it would be intriguing to consider thermal biphotons for structured illumination quantum imaging\cite{hong2017heisenberg,hong2018two}. In addition, extending our analysis and relating the breakdown of the fluctuating-speckle interpretation with spatial quantum correlations might open new avenues in the developing field of entanglement quantification and certification\cite{bavaresco2018measurements,friis2019entanglement,valencia2020high}.

\bibliography{aipsamp}
\textbf{Supplementary Material:}
The supplementary material includes a complete derivation of Eq.\ref{eq1} and the theory of thermal biphotons, detailed information on thermal biphotons under the double-Gaussian approximation and additional experimental results recovering Siegert's relation for $d\approx D$. The supplementary material can be found at https://www.scitation.org/doi/suppl/10.1063/5.0085342.

\textbf{Acknowledgments:} We acknowledge support from the Zuckerman STEM Leadership Program, the Israel Science Foundation (grant No. 2497/21), the ISF-NRF Singapore joint research program (grant No. 3538/20) and the United States-Israel Binational Science Foundation (BSF) (Grant No. 2017694).

\end{document}